\documentclass[12pt]{article}
\usepackage{amsmath}
\usepackage{amssymb}
\usepackage{amsthm}
\usepackage{array}
\usepackage{graphicx}
\usepackage{hyperref}
\usepackage{verbatim}
\usepackage{color}
\usepackage{subfig}
\usepackage{wrapfig}
\usepackage{float}
\usepackage{caption}
\usepackage[section]{placeins}
\edef\restoreparindent{\parindent=\the\parindent\relax}
\usepackage{parskip}
\restoreparindent
\usepackage[normalem]{ulem}
\usepackage{fullpage}

\graphicspath{{figures/}}

\newcommand{\be}{\begin{equation}}
	\newcommand{\ee}{\end{equation}}

\newcommand{\hD}{\widehat \Delta}
\newcommand{\hW}{\widehat W} 
\newcommand{\hR}{\widehat R}

\newcommand{\im}{\mathrm{Im}}
\newcommand{\kker}{\mathrm{Ker}}
\newcommand{\hB}{\widehat{\Box}}

\newcommand{\pg}{\Psi}
\newcommand{\eul}{\gamma_e}
\newcommand{\taubyl}{\gamma}

\newcommand{\cO}{\mathcal O}

\newcommand{\mathsym}[1]{{}}
\newcommand{\unicode}[1]{{}}

\newcommand{\diam}{\mathcal{D}}
\newcommand{\cI}{\mathcal I}
\newcommand{\cC}{\mathcal C}

\newcommand{\hP}{\hat \Phi}

\newcommand{\hwt}{\hW_\tau}
\newcommand{\sinc}{{\rm sinc}}

\newcommand{\hr}{\widehat \rho}
 
\newcommand{\wz}{\widehat{W}^{(0)}}
 
\newcommand{\cm}{\mathrm{c}_m} 
\newcommand{\sm}{\mathrm{s}_m} 
\newcommand{\mx}{\mathrm{max}}
 
\newcommand{\ssee}{\mathcal S}
\newcommand{\taup}{{\bar{\tau}}}
 
\newcommand{\Ap}{\bar{A}}
\newcommand{\Bp}{\bar{B}}
\newcommand{\psit}{\psi^{(\tau)}}
\newcommand{\psitp}{\psi^{(\taup)}}

\newcommand{\psistp}{\psi^{*(\taup)}}
\newcommand{\mtau}{M_\tau}
\newcommand{\mtaup}{M_\taup}
\newcommand{\wtau}{W_\tau}

\newcommand{\cylev}{\varrho}
\newcommand{\ccL}{\mathcal L}
\newcommand{\hwdt}{\hW_{\tau}|_{\diam_s}}

\begin{document}
	
\title{A Spacetime Calculation of \\ the Calabrese-Cardy Entanglement Entropy}

\author{Abhishek Mathur\footnote{\it abhishekmathur@rri.res.in},  Sumati Surya and Nomaan X \\{\it {\small{ Raman Research Institute, CV
    Raman Ave, Sadashivanagar, Bangalore, 560080, India}}}}

\date{}

\maketitle

\begin{abstract}
We calculate Sorkin's spacetime entanglement entropy of a Gaussian scalar field for   complementary regions in the 2d cylinder spacetime and show that it has the  Calabrese-Cardy form. We find that the cut-off  dependent term is universal  when we use  a covariant UV cut-off as in \cite{yss}. In addition, we show that  the relative size-dependent term exhibits complementarity. Its coefficient  is however {\it not}  universal and depends on the choice of pure state. It asymptotes to the universal form within a natural class of pure states.
\end{abstract}

The Calabrese-Cardy formula for the entanglement entropy (EE) of a CFT for an interval $\cI_{s}$ of length $s$  in a circle $\cC_{\ell}$ of
circumference  $\ell$ is given by
\begin{equation}
	S = \frac{c}{3} \ln\bigg(\frac{\ell}{\pi \epsilon}\bigg)+ \frac{c}{3} \ln(\sin(\alpha
	\pi))+c_1
	\label{cc.eq}
\end{equation}  
where $\alpha={s}/{\ell}$, $c$ is the CFT central charge, $\epsilon$ is a UV cut-off and $c_1$ is a non-universal
constant \cite{cc}.  This formula has been shown to apply to a diverse range of two dimensional systems which fall 
within the same universality class,  including a geometric realisation by  Ryu and Takayanagi \cite{ryu2006holographic} and others \cite{headrick}. Entanglement entropy (EE) was first proposed in
\cite{bkls}  as a possible contributor to black hole entropy. Hence understanding Eqn.~(\ref{cc.eq}) from a spacetime
perspective is of broad interest. 

As a follow up to their earlier work, Calabrese and Cardy studied
the unitary time evolution of the EE for  an interval $\cI_s$
inside  a larger interval $\cI \supset
\cI_s$. Starting with a pure state, which is an eigenstate of a "pre-quench" Hamiltonian, and then quenching the system at $t=0$,  they used  path  integral techniques  to show that the EE increases with time.
It then saturates  after the ``light-crossing'' time,   in keeping with causality \cite{evolcc}. This corresponds to the ``time''  required for the domain of dependence of $\cI_s$ to be fully
defined. Seeking out a covariant formulation of EE  is therefore of interest both to understanding 
the results of \cite{evolcc}  in a spacetime language as
well as more generally  in QFT and quantum gravity.  Such a formulation is moreover in keeping with the broader framework of   AQFT, where observables  are
associated with spacetime  regions rather than spatial hypersurfaces \cite{fewster2020algebraic}.

In \cite{ssee}  Sorkin proposed a  spacetime formula for  the EE of a Gaussian scalar field $\Phi$  in a globally hyperbolic subregion $\cO$ of a globally hyperbolic spacetime $(M,g)$, with respect to its causal complement $\cO^c$. It  uses the restriction of the Wightmann function $W(x,x')$ in $M$ to  $\cO$, and the Pauli-Jordan function $ i \Delta(x,x')$  which appears in the  Peierl's spacetime commutation relation  $[\hP(x),\hP(x')] = i \Delta(x,x').$ 

Sorkin's spacetime EE (SSEE)  of $\cO$ with respect to $\cO^c$ is 
\begin{equation}
	\ssee = \sum_\mu \mu \ln (|\mu|),   \quad   \hW|_{\cO} \circ \chi = \mu (i \hD) \circ \chi, 
	\label{ssee.eq}
\end{equation}
where $\chi \not\in \kker( \hD)$ and where  
\begin{equation}
	A\circ v (x) \equiv \int_{\cO} dV_{x'} A(x,x') v(x').  
\end{equation}
It is motivated by the finite system Wightmann function for a Gaussian state which is a direct sum of identical systems with two degrees of freedom \cite{ssee}. The SSEE formula generalises the calculation of EE for  a state at a given time to that associated with  a 
spacetime region. 

\begin{wrapfigure}{R}{0.5\textwidth}
	\centering{\includegraphics[height=6.6cm]{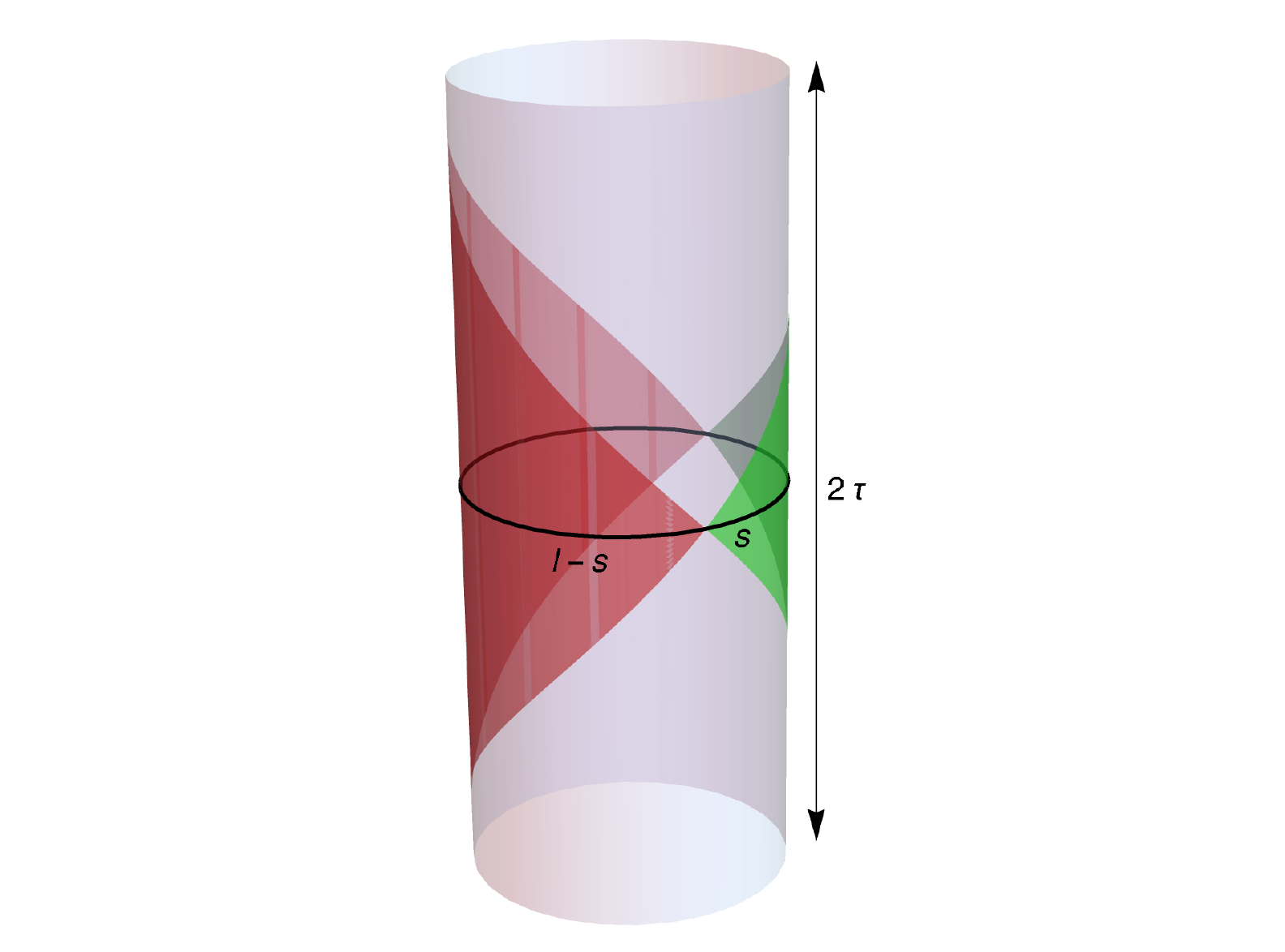}}
	\caption{The spacetime analogues of $\cI_{s}, \cI_{\ell-s} \subset \cC_\ell$ are their domains of dependence  $\diam_s$ and $\diam_{\ell-s}$ in $(M,g)$ shown in green and red respectively.}
	\label{cyl.fig} 
\end{wrapfigure}  

In \cite{yss} the SSEE for nested causal diamonds $\diam_s \subset \diam_S$ was shown to yield the first, cut-off dependent term of Eqn.~(\ref{cc.eq})  with $c=1$ when $s<<S$.  Since $\diam_s$ is the domain of dependence of $I_s$, this is the natural spacetime analogue of $\cI_s \subset \cI_S$. In this work we calculate the SSEE for the spacetime analogue of $\cI_s \subset \cC_\ell$ for finite $\ell$ and additionally, find the same $\alpha$-dependence as Eqn.~(\ref{cc.eq}), thus explicitly demonstrating  complementarity. 
A natural spacetime analogue of $\cC_{\ell} $ is its (zero momentum) Cauchy completion, which is the $d=2$ cylindrical spacetime $(M, g)$ with $ds^2=-dt^2 + dx^2, \,  x+\ell \sim x$. The domains of dependence of $\cI_{s} $ and its complement $\cI_{\ell-s}$ in $(M,g)$ are the  causal diamonds $\diam_{s}$ and $\diam_ {\ell -s}$ respectively, as shown in Fig \ref{cyl.fig}.

In what follows we use a mixture of analytical and numerical methods to solve the SSEE eigenvalue problem. 

We will find it convenient to work with the Sorkin-Johnston (SJ)  formulation \cite{Sorkin:2017fcp,fv,Brum:2013bia,afshordi2012distinguished,Afshordi:2012ez}, where the SJ spectrum  provides the required (covariant)  UV cut-off with which to calculate  $\ssee$, as was done in \cite{yss}. For a {compact} globally hyperbolic region $(M,g)$ of a spacetime it follows from $\kker(\hB) = \im(i\hD)$ \cite{wald1994quantum} that the eigenmodes of the integral Hermitian operator $i\hD$ provide a  covariant orthonormal basis (the SJ modes) with respect to  the $\ccL^2$ norm on $(M,g)$ \cite{fewster2020algebraic}. The  SJ vacuum or Wightmann function is given by the positive part of $i\hD$.  Since the SJ spectrum is covariant so is a UV cut-off in this basis.   

For our calculation of  $\ssee$ we will use the SJ vacuum $\wtau$ for a free massless scalar field in  a  slab  $(\mtau, g)$ of
height $2 \tau$  in the cylinder  spacetime \cite{fv}, and its
restriction to $\diam_s \subset \mtau$,  
\begin{equation}
	\wtau(x,t; x',t') = \sum_{m\in \mathbb Z} \cylev_{m} \psi_m(x,t) \psi^*_m(x',t'),  \label{cylsj.eq}
\end{equation} 
where $\{ \psi_m, \cylev_m\}$ are the $\ccL^2$ normalised positive frequency SJ eigenmodes and eigenvalues in $\mtau$
\cite{fv}: 
\begin{align} 
	&\psi_m(x,t)\! = \!\biggl(\!\frac{\!(1 \!- \!\zeta_m\!)}{2\sqrt{2\ell}\cm}\!e^{i \!\frac{2 \pi |m| t}{\ell}} \! +  \!\frac{\!(\!1 \!+ \!\zeta_m\!)}{2\sqrt{2\ell}\cm} \!e^{-i \!\frac{2 \pi |m|t}{\ell}}\!\biggr)e^{i \!\frac{2 \pi m x}{\ell}} \nonumber  \\ 
	&\cylev_m  =   \ell
	\frac{\sm\cm }{2 \pi |m|}, \,    \zeta_m=\frac{\cm}{\sm}, \,  \taubyl=\frac{2\tau}{\ell}, \quad m \in \mathbb Z, \nonumber\\
	&\cm^2\! = \! \tau\left(1+\sinc(2 |m| \pi \taubyl)\right),  \sm^2\! =\! \tau\left(1-\sinc( 2 |m| \pi \taubyl)\right).
	\label{cylsjef.eq}
\end{align} 

The $m=0$ ``zero mode''  in particular takes the  form 
\begin{equation} 
	\psi_0(t)= \frac{1}{2\sqrt{\tau l}}\left(1 - i \frac{\sqrt{3}}{\tau}t\right),  \quad \cylev_0=
	\frac{2}{\sqrt{3}}\tau^2. 
\end{equation}
Unlike  the standard vacuum on the cylinder, $\wtau$ is $\tau$-dependent. Each  $\wtau$  can however be viewed as  a pure (non-vacuum) state in $\mtaup$  for any $\taup >
\tau$, as we will later show. To accommodate both  $\diam_{s}$ and 
$\diam_{\ell-s}$ in our calculations,  we require $ 2 \tau  \geq s, \ell-s$.

The  SJ modes in $\diam_{s}$ are naturally expressed in terms of
the light cone coordinates $u=\frac{1}{\sqrt{2}}(t-x), v=\frac{1}{\sqrt{2}}(t+x)$ and   come in the two
mutually 
orthogonal series   \cite{jthesis}
\begin{align}
	f_k \!&\!=\! e^{-iku}\!-\!e^{-ikv},  \quad  k = 2 \sqrt{2} n \pi/s \nonumber \\ 
	g_\kappa \!&\!=\!  e^{-i\kappa u}\!+\!e^{-i\kappa v}\!-\!2\cos\bigg(\frac{\kappa s}{2 \sqrt{2}}\bigg),\quad \tan\bigg(\frac{\kappa s
	}{2\sqrt{2}}\bigg)=\frac{\kappa s}{\sqrt{2}}\, \, 
	\label{diamsj.eq}
\end{align}
with eigenvalues $\lambda_{k}= \dfrac{s}{2\sqrt{2}k}$ and $ \lambda_{\kappa}=
\dfrac{s}{2 \sqrt{2}\kappa}$, respectively, and with $\ccL^2$ norm in  $\diam_{s}$ 
\begin{equation}
	{||f_k||^2 =s^2  , \quad ||g_\kappa||^2 = s^2\left(1-2\cos^2\left(\frac{\kappa s}{2\sqrt{2}}\right)\right)}. 
\end{equation} 
Since $i \hD $  is diagonal in this basis we will use it to transform Eqn.~(\ref{ssee.eq}) to the matrix form 
\begin{equation} 
	\hW_\tau|_{\diam_s} X   =  \mu \Lambda X,  
\end{equation} 
where $\Lambda$ is the diagonal matrix $\{\lambda_k, \lambda_\kappa \}$.  
For $X \not\in
\kker(i\hD)$, we can invert this to suggestively write 
\begin{equation}
	\hr X =  \Lambda^{-1} \hW_\tau|_{\diam_s} X     =\mu X, 
\end{equation} 
so that  $\ssee$ can be viewed as  the
von-Neumann entropy of $\hr$.  The spectrum of $\hr$ is  unbounded and hence needs 
a UV cut-off. As in \cite{yss}
we use the  covariant UV-cut off with respect to  the SJ spectrum $\{\lambda_k,\lambda_\kappa\}$. For large $\kappa$  the condition $\tan({\kappa s}/{2\sqrt{2}})=\kappa s/\sqrt{2}$ can be approximated by $\kappa \sim\sqrt{2}(2n+1)\pi/s$, so that a consistent choice of cut-off for both sets of eigenvalues is $\epsilon=k_\mx^{-1}= s/(2 \sqrt{2} n_\mx \pi)$.  We also  need to ensure that this same cut-off is used in the causal complement, i.e., $k_\mx=2 \sqrt{2} n'_\mx \pi/(\ell-s)$, where $n'$ denotes the quantum number for the SJ spectrum in $\diam_{\ell -s}$, 
so that $\epsilon \!=\!\dfrac{\ell \alpha}{2\sqrt{2}\pi n_\mx}\!=\! \dfrac{\ell (1-\alpha)}{2\sqrt{2}\pi n_\mx'}$.

We expand the SJ  modes in $\mtau$  in terms of those in $\diam_s$ to obtain the  non-zero matrix elements for $\hwdt$ for general $\alpha, \gamma$. Suppressing  the $\tau, \diam_s$ labels, these are    
\begin{align}
	\hW_{kk'}\!&=\! \frac{s^4}{32\pi} \sum_{m> 0}\frac{1}{|m|\zeta_m} \biggl(\eta^-_m\sinc(x_+)-\eta^+_m \sinc(x_-)\biggr) \times \biggl(\eta_m^-\sinc(x_+') - \!\eta_m^+ \sinc(x_-') \biggr) \nonumber \\
	\hW_{\kappa\kappa'}\!&=\! \frac{s^4}{32\pi} \sum_{m> 0}\! \frac{1}{|m|\zeta_m} 
	\biggl(\eta_m^-\sinc(z_+) +\eta_m^+ \sinc(z_-)\biggr) \times \biggl(\eta_m^-\sinc(z_+') + \eta_m^+ \sinc(z_-')
	\biggr) \nonumber\\
	&\hspace{8cm}+\wz_{\kappa\kappa'}  \label{wmatrix.eq} 
\end{align}
where $x_\pm=(n\pm\alpha m)\pi$,  $x_\pm'=(n'\pm \alpha m)\pi$,  $z_\pm=\kappa s/2\sqrt{2}\pm \alpha m\pi$,   $z_\pm'=\kappa' s/2\sqrt{2}\pm \alpha m\pi$,  and  the contribution from the zero mode is
\begin{align} 
	\wz_{\kappa\kappa'} &= \frac{s^4}{2 \sqrt{3}} \frac{\tau}{\ell} \cos(\kappa s/(2\sqrt{2})) \cos(\kappa' s/(2\sqrt{2}))\nonumber \\ 
	& \times \biggl(1+ \sqrt{\frac{3}{2}} \frac{1}{\kappa \tau} \biggr)   \biggl(1+ \sqrt{\frac{3}{2}} \frac{1}{\kappa'
		\tau}\biggr).  \label{wzeromode.eq} 
\end{align}

Our strategy is to  construct $\hr\,$ from these matrix elements and to solve for its eigenvalues using a numerical matrix solver. However, each matrix elements in Eqn.~(\ref{wmatrix.eq}) is an infinite sum over the quantum number $m$ and hence not amenable to explicit calculation. 
We therefore need to find a closed form expression for the above matrix elements. 

We notice that when $\taubyl$ takes  half-integer values (for which the SJ vacuum  Hadamard \cite{fv}), $\zeta_m=1$ for $ m \neq 0$, which leads to a considerable simplification. Further, let   $\alpha$ be rational,  so that we can write  $\alpha=\frac{p}{q}$, with   $p, q \in \mathbb Z,$ and $  p, q  >0$  being relatively prime. For these choices of $\alpha$ and $\gamma$, the infinite sums of Eqn.~(\ref{wmatrix.eq}) reduce to the following finite sums over Polygamma functions $\pg(x)$ and $\pg^{(1)}(x)$
	
	\begin{align}
		\hW_{kk'}& =  \frac{ s^4}{8 \pi n}  \Biggl[ \delta_{n,n'} \biggl( \alpha  \Theta(n) \sum_m
		\delta_{n,m\alpha} + 
		\frac{1}{\pi^2 \alpha q^2 n} \sum_{r=1}^{q-1} \sin^2(r\alpha \pi) \biggl[-\alpha q  \pg\!\Bigl(\frac{r}{q}\Bigr) + \alpha q
		\pg\!\Bigl(\frac{\alpha r -n}{\alpha q}\Bigr)\nonumber\\
		&+ n \pg^{(1)}\!\Bigl(\frac{\alpha r  -n}{\alpha q}\Bigr)\biggr]\biggr) 
		+ (1-\delta_{n,n'}) \frac{(-1)^{n+n'}}{\pi^2 n'(n-n') q}  \sum_{r=1}^{q-1} \sin^2(r \alpha \pi) \biggl[ (n'-n)
		\pg\!\Bigl(\frac{r}{q}\Bigr)\nonumber\\
		& - n' \pg\!\Bigl(\frac{\alpha r -n}{\alpha q}\Bigr) + n \pg\!\Bigl(\frac{\alpha r
			-n'}{\alpha q} \Bigr)\biggr] \Biggr]  \nonumber \\
		\hW_{\kappa\kappa'} &=s^4\cos\left(\frac{\kappa
			s}{2\sqrt{2}}\right)\cos\left(\frac{\kappa's}{2\sqrt{2}}\right)\Biggl[\frac{\tau}{2\sqrt{3}
			\ell}\biggl(1+\sqrt{\frac{3}{2}} \frac{1}{\tau\kappa}\biggr) \biggl(1+\sqrt{\frac{3}{2}} \frac{1}{\tau\kappa'}\biggr)\nonumber\\
		& +  \delta_{\kappa,\kappa'} \frac{1}{\alpha q^2s^2\kappa^2\pi^2}\biggr(
		\sum_{r=1}^{q-1}\Omega(\kappa,\kappa',\alpha,r)\biggl[\alpha q \pi\biggl(\pg\!\Bigl(\frac{r}{q}-\frac{\kappa
			s}{\eta}\Bigr) - \pg\!\Bigl(\frac{r}{q}\Bigr)\biggr) +\frac{\kappa
			s}{2\sqrt{2}}\pg^{(1)}\!\Bigl(\frac{r}{q}-\frac{\kappa s}{\eta}\Bigr)\biggr] \nonumber\\
		& +   \frac{s^2\kappa\kappa'}{2} \biggl[ \alpha q  \pi\biggl(\eul + \pg\!\Bigl(1-\frac{\kappa s}{\eta}\Bigr)\biggr) + \frac{\kappa
			s}{2 \sqrt{2}}\pg^{(1)}\!\Bigl(1-\frac{\kappa s}{\eta}\Bigr)\biggr] \biggr)  +  (1-\delta_{\kappa,\kappa'}) \frac{1}{s^2 q \kappa\kappa' (\kappa-\kappa')}\nonumber\\
		&\times \biggl( 
		\sum_{r=1}^{q-1}\Omega(\kappa,\kappa',\alpha,r) \biggl[\kappa\pg\!\Bigl(\frac{r}{q}-\frac{\kappa's}{\eta}\Bigr) - \kappa'\pg\!\Bigl(\frac{r}{q}-\frac{\kappa s}{\eta}\Bigr) - (\kappa-\kappa')\pg\!\Bigl(\frac{r}{q}\Bigr)\biggr]\nonumber\\
		&+ \frac{s^2\kappa\kappa'}{2} \biggl[\eul(\kappa-\kappa') + \kappa\pg\!\Bigl(1-\frac{\kappa's}{\eta}\Bigr)
		-\kappa'\pg\!\Bigl(1-\frac{\kappa s}{\eta}\Bigr) \biggr] \biggr)\Biggr], \qquad \eta=2\sqrt{2} \alpha q
		\pi  \label{wmatrixgg.eq} 
	\end{align} 

where $\eul$ represents the Euler-Mascheroni constant and
\begin{align}
	\Omega(\kappa,\kappa',\alpha,r)&=\kappa\kappa' \frac{s^2}{2} \cos^2(\alpha r\pi) + \sin^2(\alpha r\pi)\nonumber \\
	&-(\kappa+\kappa')\frac{s}{2\sqrt{2}}\sin(2\alpha r\pi).
\end{align}

We are now in a position to solve for the eigenvalues of $\hr$ using Mathematica's numerical eigenvalue solver. We consider a range of values of $\alpha, \gamma$ and the cut-off $n_\mx/\alpha$ given in the table below. 

\begin{table}[h]
	\begin{center}
		{\renewcommand{\arraystretch}{1.2}
			\begin{tabular}{|c|c|}
				\hline
				\rule{0pt}{3ex}$\alpha$&
				$\frac{1}{10},\frac{1}{5},\frac{1}{4},\frac{1}{3},\frac{1}{2},\frac{2}{3},\frac{3}{4},\frac{4}{5},\frac{9}{10}$ 
				\\ [1ex]\hline \rule{0pt}{3ex}
				$\gamma$& $1,2,4,6,8,16,21.5,32,40.3,100,200,1000,2000$\\  [1ex] \hline \rule{0pt}{3ex}
				$\frac{n_\mx}{\alpha}$& $1000,1200,1400,1600,1800,2000,2200,2400,2600$\\[1ex] 
				\hline
			\end{tabular}
		}
	\end{center} 
\end{table}
In the list of $\gamma$ values, we have also included the specific non-half-integer value  of
$\gamma =40.3 $ for which $\zeta_m \sim 1$ even for $m =1$.   In general, we note that   $\zeta_m \sim 1$ for  $m
>> \gamma^{-1}$. The error coming from small $m$ terms
has been explicitly calculated in this case as a function of $m$ and seen to be small.
For the special case $\alpha=0$,  $\ssee$ is trivially zero, while for
$\alpha=1$, the domain of dependence of $\cC_\ell$ is no longer a causal diamond, but all of $\mtau$. Since
$\hwt$ is the SJ vacuum  and therefore pure, $\ssee=0$.

\begin{figure}[!ht]
	\centering
	\includegraphics[scale=0.75]{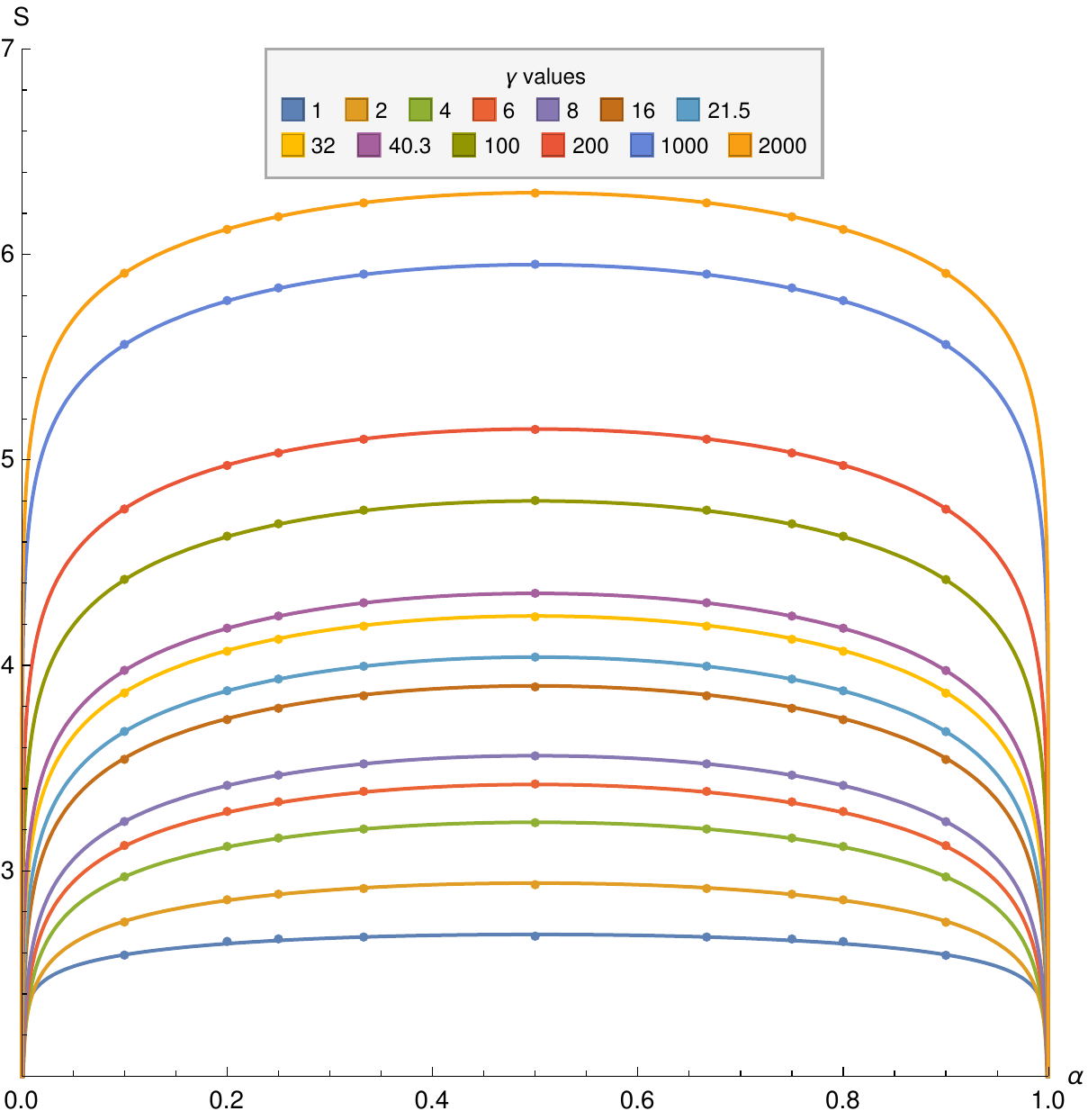}
	\caption{$\ssee$ vs $\alpha$ for different $\gamma$ fitted to 
		$\ssee=a\log(\sin(\pi\alpha))+b$, with $\frac{n_\mx}{\alpha}=2600$.
	}
	\label{salpha.fig} 
\end{figure}

\begin{figure}[!ht]
	\centering{\includegraphics[scale=0.75]{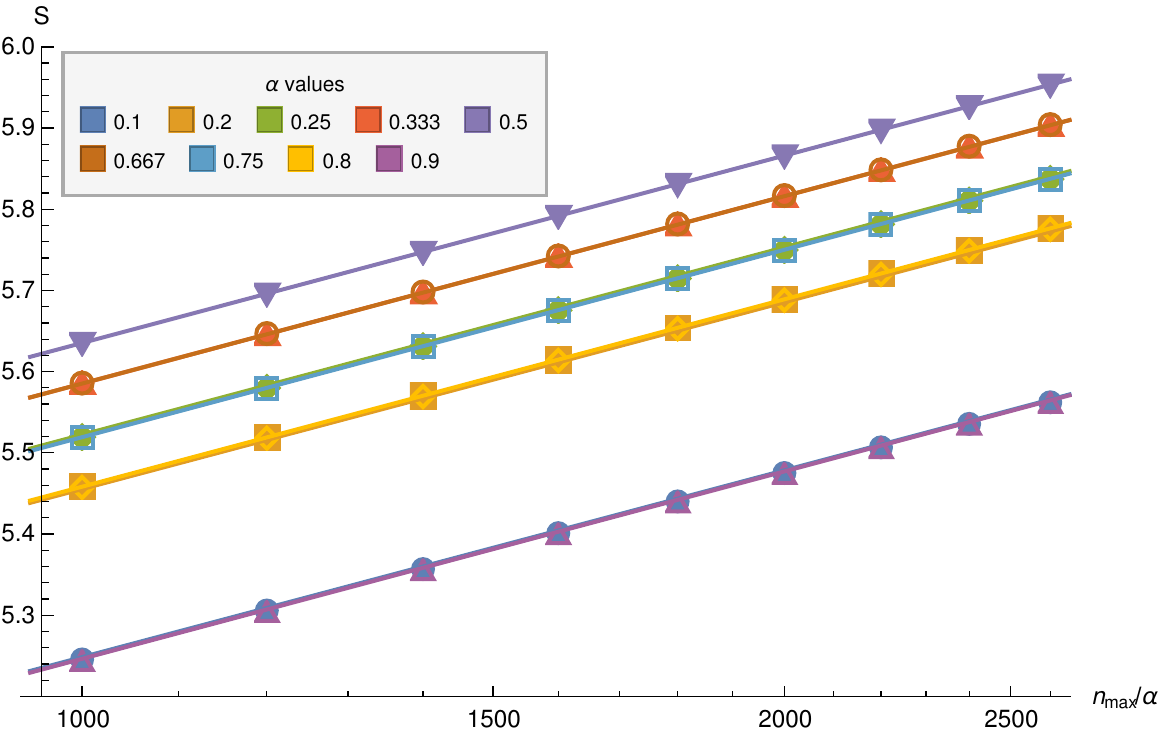}}
	\caption{Log-linear plot of  $\ssee$ vs $\frac{n_\mx}{\alpha}$ for different $\alpha$  fitted to
		$\ssee=a\log\,(n_\mx/\alpha)+b$ for $\gamma=1000$. 
	} \label{Compl.fig} 
\end{figure}

Fig.~\ref{salpha.fig}  shows the results of simulations for these  various $\alpha$ and $\gamma$ values, for a fixed choice of cut-off $n_\mx/\alpha=2600$. It is already clear that $\ssee$ satisfies 
complementarity. This is much more explicit in Fig.~\ref{Compl.fig}, where we vary over the cut-off.  
Our numerical results suggest that  $\ssee$ takes the general  form
\begin{equation}
	\ssee  = \frac{c(\gamma)}{3} \ln\bigg(\frac{\ell}{\pi \epsilon}\bigg)+ {f(\gamma)}\ln(\sin(\alpha
	\pi))+c_1(\gamma).  \label{ourresult.eq}
\end{equation} 
Using the best-fit curves in Figs.~\ref{Compl.fig}-\ref{conegamma.fig}, and the associated data in the appendix, we find that  $c(\gamma) \sim 1$ and 
\begin{align} 
	f(\gamma) & \sim  0.33+a/\gamma+b/\gamma^2 \nonumber \\
	c_1(\gamma) &\sim  a'\log\gamma+b'.
	\label{cfcone.eq}
\end{align}

\begin{figure}[h!]
	\centering
	\includegraphics[scale=0.8]{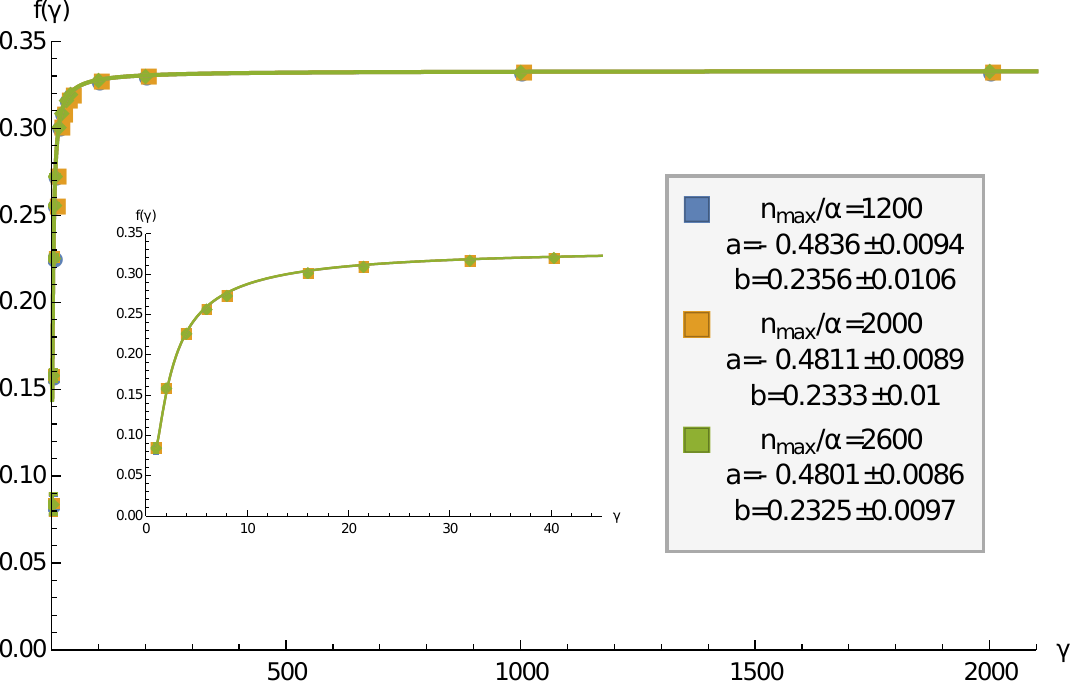}
	\caption{A plot of $f(\gamma)$ vs. $\gamma$ for different values of $n_\mx/\alpha$, fitted to 
		$0.33+a/\gamma+b/\gamma^2$.  
		The inset figure shows the smaller $\gamma$ values.} \label{fgamma.fig} 
\end{figure}
\begin{figure}[!h]
	\centering
	\includegraphics[scale=0.8]{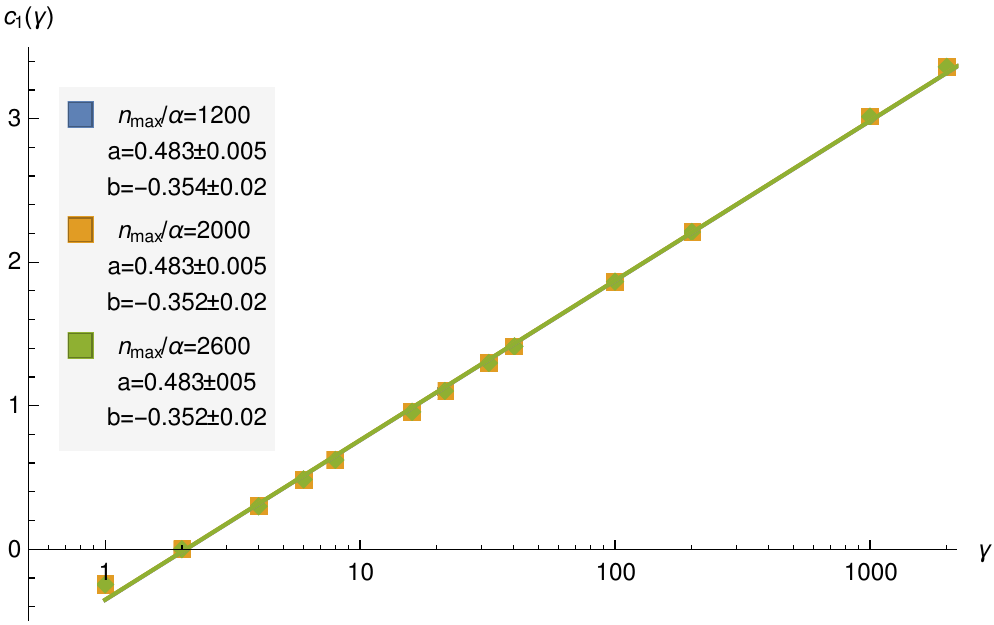}
	\caption{A log-linear plot of $c_1(\gamma)$ vs $\gamma$ for different values of $n_\mx/\alpha$ 
		fitted to  $a\log\gamma+b$. }\label{conegamma.fig}
\end{figure}

Thus, the first term of Eqn.~(\ref{cc.eq}) is reproduced for any choice
of $\alpha, \gamma$.  
This  generalises the results of \cite{yss}  where this was shown in the limit of $\alpha \ll 1$. The dependence on $\alpha$, i.e., the second term of  Eqn.~(\ref{cc.eq}) is also reproduced and hence exhibits complementarity for any $\alpha$ 
(see Fig.~\ref{salpha.fig}, ~\ref{Compl.fig}).  Its  coefficient however is {\it not}
universal and depends on $\gamma$ as shown in Fig.~\ref{fgamma.fig}.  However, as $\gamma >>1$,  
$f(\gamma) $ does asymptote to the universal value $1/3$. Finally,  the non-universal constant
$c_1(\gamma)$  diverges logarithmically  with $\gamma$ as shown in 
Fig.~\ref{conegamma.fig}. This can be traced to the IR divergence in the zero modes of the massless theory. 

The behaviour of $f(\gamma)$ can be viewed as a
dependence on the choice of the pure state $W_\tau$ in $\mtaup$, for $\mtaup \supset \mtau$. From  Eqn.~(\ref{cylsj.eq}) we see that $\hW_\tau$ is a state in $\mtaup$, i.e., 
$\hW_\tau=\hR_\tau+ i \hD/2$, where $\hR_\tau$ is real and symmetric.  

Expanding $i\hD$ in    the SJ modes $\{ \psitp_m \}$ of $\mtaup$ and $\hW_\tau$  in $\{ \psit_m\}$, and inserting in Eqn.~(\ref{ssee.eq}) 
we see that term by term 
\be
\cylev^{\tau}_m\psit_m(x,t) A_m= \mu
\cylev^\taup_m[\psitp_m(t,x) \Ap_m\ \!-\! \psistp_m(t,x)\Bp_m],  \nonumber 
\ee
where $A_m \!=\! (\!\psit_m,\chi\!)_\taup$, $\Ap_m\!=\!(\!\psitp_m,\chi\!)_\taup$,  $\Bp_m\!=\!(\psistp_m,\chi\!)_\taup$ and $(.,.)_\taup$ is the $\ccL^2$ inner product in $\mtaup$. 
Expanding  $\psit_m=a_m\psitp_m + b_m\psistp_m$,   $a_m=\frac{\sm^{\tau'}}{2 \cm^{\tau}} (\zeta_m^ {(\tau')}
+\zeta_m^ {(\tau)}) $,  $b_m=\frac{\sm^{\tau'}}{2 \cm^{\tau}} (\zeta_m^ {(\tau')}
-\zeta_m^ {(\tau)}) $  this simplifies to
\begin{align}
	\cylev^{\tau}_m a_m\!\left(\! a_m \Ap_m+ b_m \Bp_m\!\right) 
	\!&=\! \mu\cylev^{\taup}_m \Ap_m \nonumber\\
	\cylev^{\tau}_m b_m\!\left(\!a_m \Ap_m + b_m \Bp_m\!\right)\!&=\!-\mu\cylev^{\taup}_m \Bp_m.                               \end{align}
The solutions for this are  either $a_m \Ap_m + b_m \Bp_m =
0 \Rightarrow \mu=0$, or 
$
a_m \Bp_m + b_m \Ap_m = 0 \Rightarrow \mu
=\frac{\cylev^{\tau}_m}{\cylev^{\tau'}_m}\left(a_m^2-b_m^2\right)=1,   
$
which means that $\hW_\tau$ is a pure state in $\mtaup$. Thus $f(\gamma)$ can be viewed as the dependence  on the choice of  pure state in  $\mtaup$ for $\mtau \subset\mtaup$. 

We end with some remarks. While we have demonstrated complementarity for certain rational values of $\alpha$,  an analytic  demonstration using Eqn.~(\ref{wmatrixgg.eq}) seems non-trivial, in part because the UV regulated matrices $\hr_\alpha$ and $\hr_{1-\alpha}$  are of different
dimensions.  
Conversely, complementarity implies that if $n_\mx>n_\mx'$,   $\hr_\alpha=\hr_{1-\alpha}
\oplus \mathbf{1}_{N} \oplus \mathbf{0}_{N}$, where $\mathbf{0}$ is the zero matrix and $N=(n_\mx-n_\mx')/2$.

In our computations we find that the eigenvalues of $\hr$ (which always come in pairs $(\mu,1-\mu)$) exhibit the  surprising feature
that all but one pair hovers around the values $0$ and $1$, 
thus contributing most significantly to $\ssee$. Indeed, the  $\ssee$ calculated using the largest few pairs of eigenvalues accounts for most of the entropy (see appendix).

Finally, it would be interesting to calculate the non-zero mass case which is  IR divergence free. While the small mass approximation of the SJ modes in $\diam_s$ is known \cite{ms}, the challenge will be to obtain closed form expressions for the matrix elements of $\hW$ as we have done.

\section*{Acknowledgements}  SS is supported in part by a Visiting Fellowship at the Perimeter Institute. Research at Perimeter Institute is supported in part by the Government of Canada through the  Department  of  Innovation,  Science  and  Economic  Development  Canada  and  by  the Province of Ontario through the Ministry of Colleges and Universities.

\section*{Appendix: Supporting Data}

In this appendix we present some plots with additional data which were used to compute the coefficients $c(\gamma)$, $f(\gamma)$ and $c_1(\gamma)$ in the Entanglement Entropy.

\begin{figure*}
	\centering
	\subfloat[$n_{max}/\alpha=1200$]{\includegraphics[height=5.4cm]{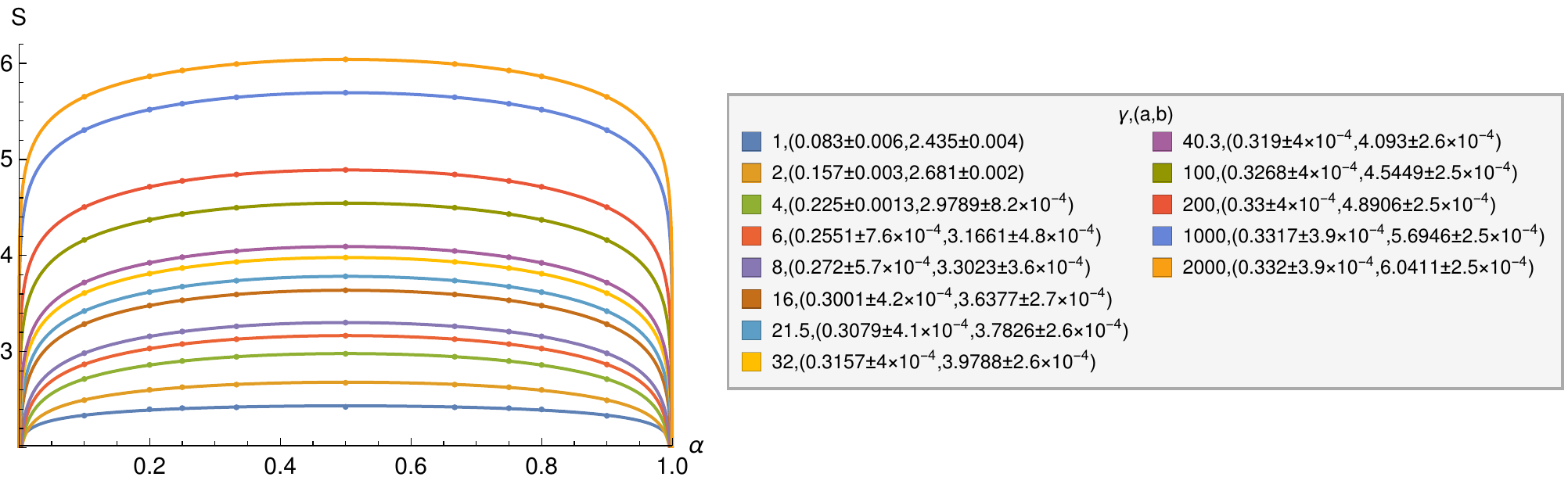}}\hskip 0.1in
	\subfloat[$n_{max}/\alpha=2000$]{\includegraphics[height=5.4cm]{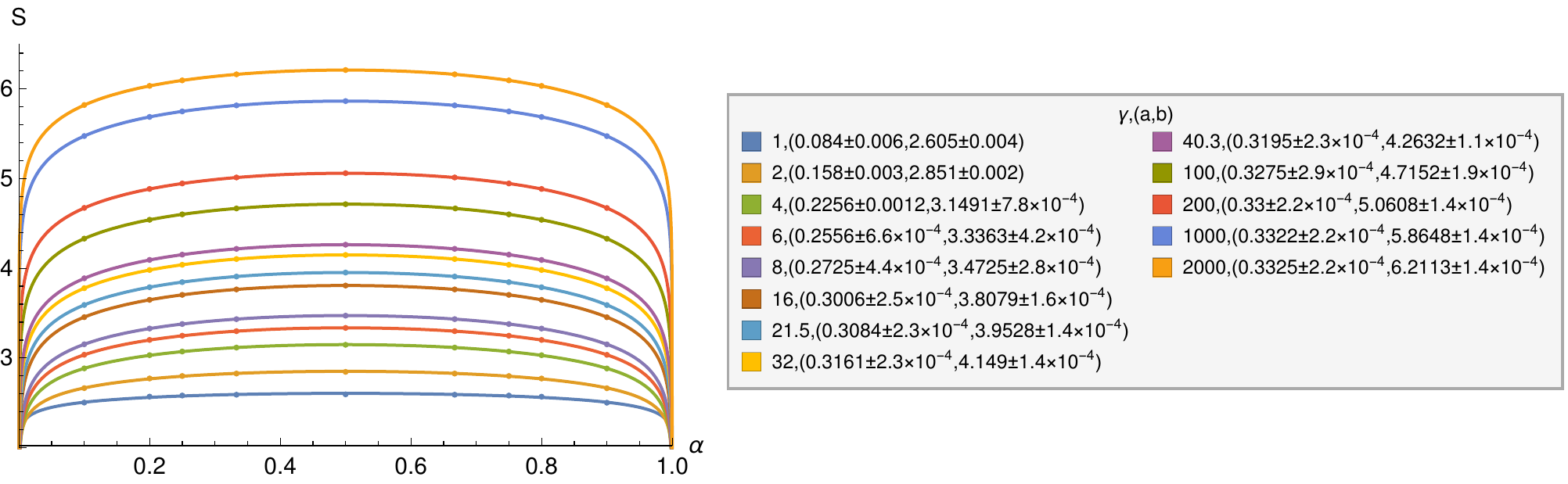}}\hskip 0.1in
	\subfloat[$n_{max}/\alpha=2600$]{\includegraphics[height=5.4cm]{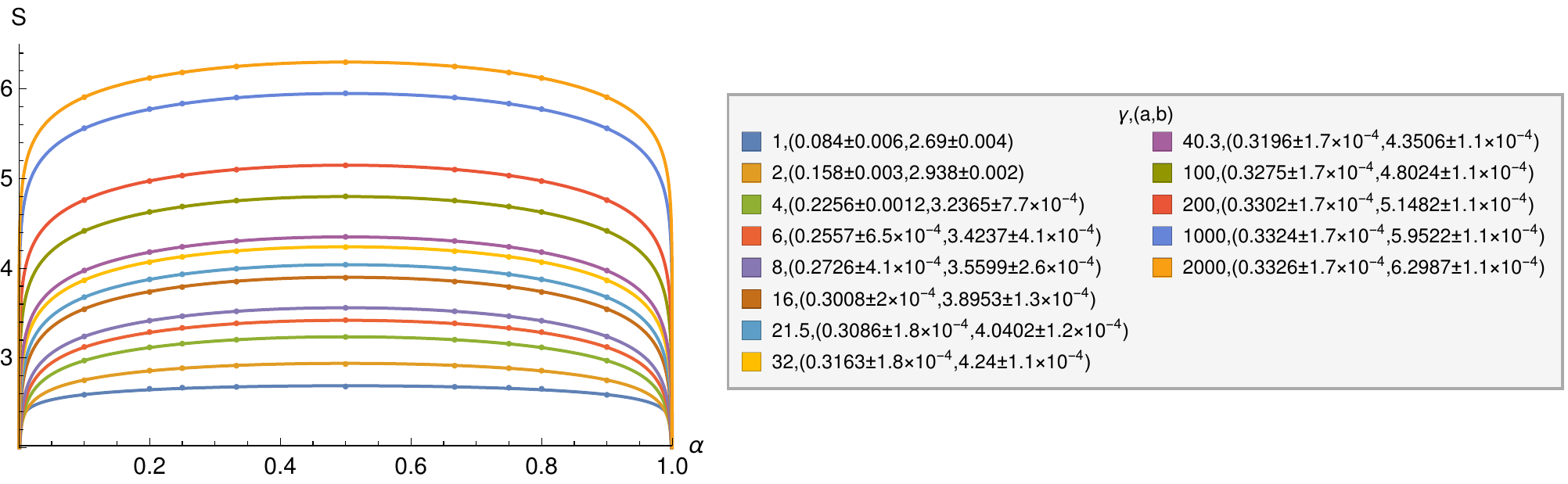}}
	\caption{$\ssee$ vs $\alpha$ for different $\gamma$ with $n_\mx/\alpha=1200,\, 2000$ and $2600$ fitted to 
		$\ssee=a\log(\sin(\pi\alpha))+b$. The fit parameters are shown in the table. }
	\label{Allsalpha.fig}
\end{figure*}

Fig.~\ref{Allsalpha.fig} shows the dependence of $\ssee$ on $\alpha$ for different values of $\gamma$ and with three different values of $n_{max}/\alpha$ $(1200,2000\text{ and }2600)$. The SSEE can be fitted to the form $\ssee = a_1\log(\sin(\alpha\pi))+b_1$ where the coefficient $a_1$ corresponds to $f(\gamma)$ in Eqn.~\eqref{ourresult.eq}. The values of $a_1$ and $b_1$ along with their errors are given in the tables in Fig.~\ref{Allsalpha.fig}. $a_1$ and therefore $f(\gamma)$ can be seen to be independent of $n_{max}/\alpha$. It is however dependent on $\gamma$ and asymptotes to the universal value of $1/3$ in the Calaberse-Cardy formula for $\gamma>>1$. We fit $f(\gamma)$ values to the form
\begin{equation}
f(\gamma)=0.33 + a_2/\gamma + b_2/\gamma^2
\end{equation}
and find the $a_2\approx-0.48$ and $b_2\approx0.23$ with the error given in the tables of Fig.~\ref{fgamma.fig}.

\begin{figure*}
	\centering
	\subfloat[$\gamma=16$]{\includegraphics[height=6cm]{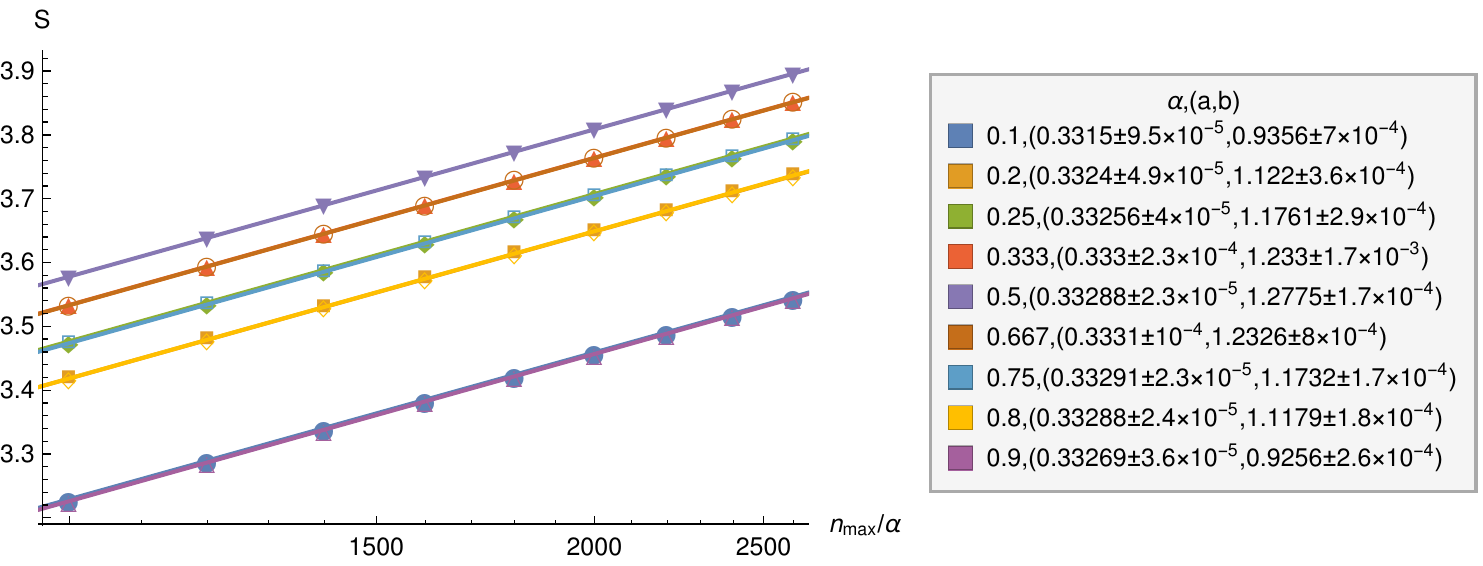}}\\
	\subfloat[$\gamma=200$]{\includegraphics[height=6cm]{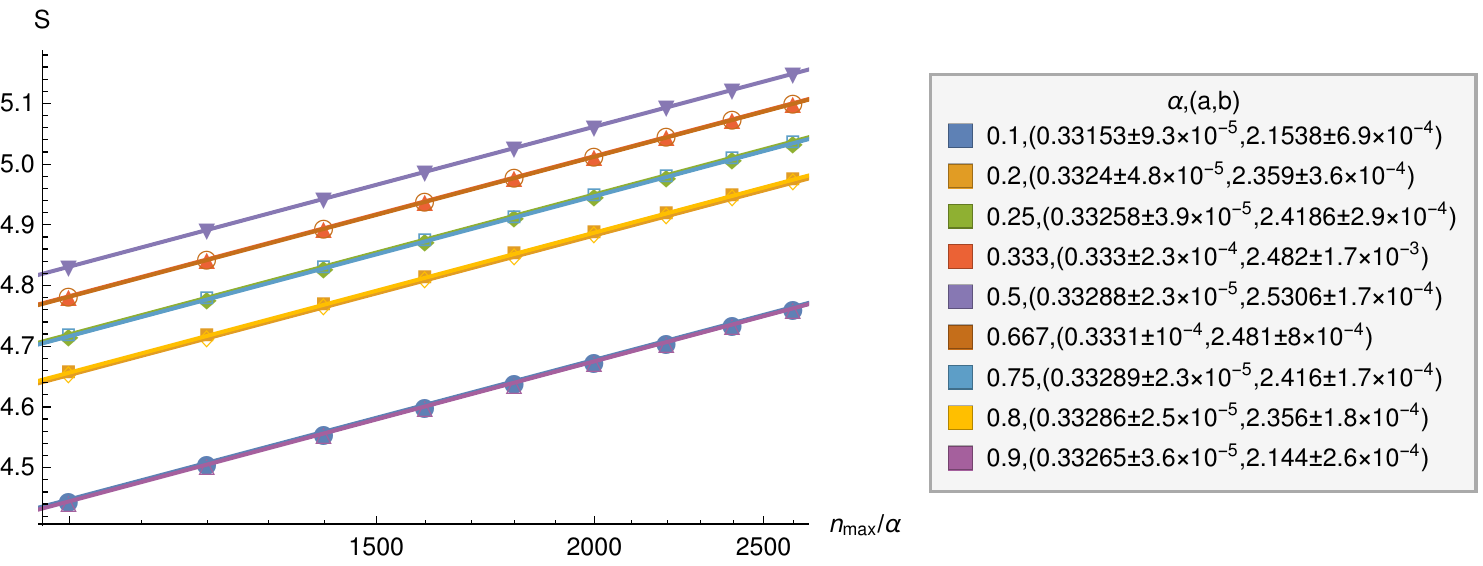}}\\
	\subfloat[$\gamma=1000$]{\includegraphics[height=6cm]{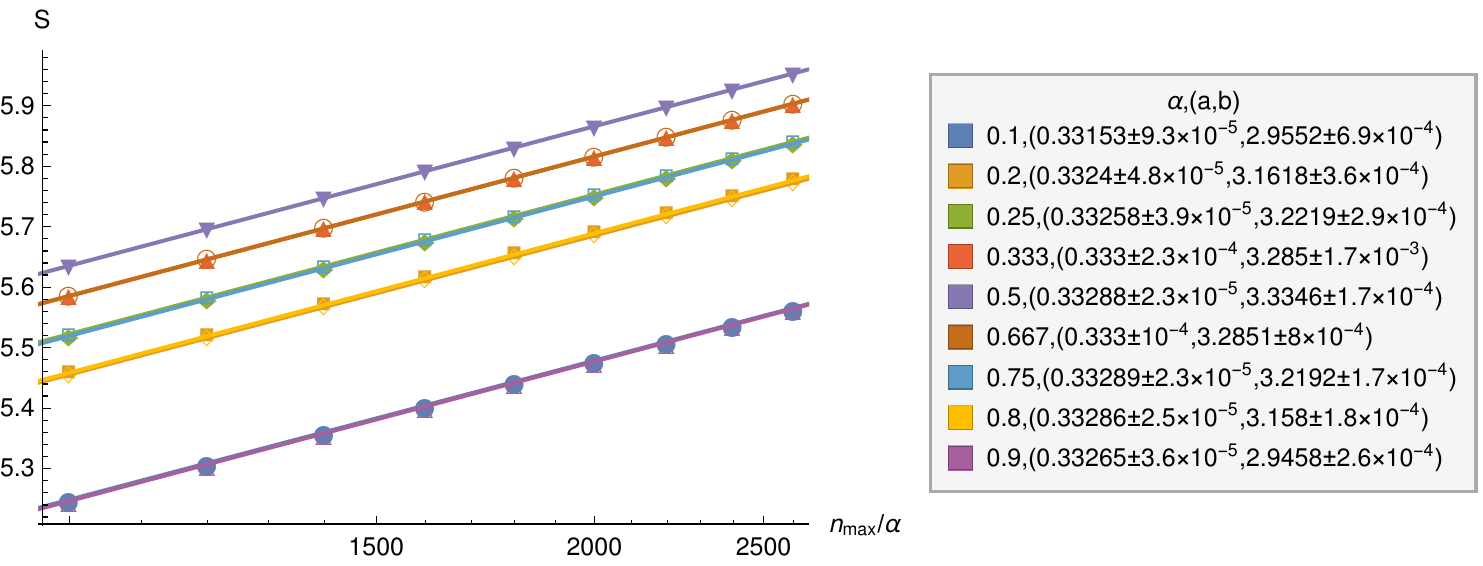}}
	\caption{A log-linear plot of  $\ssee$ vs $n_\mx/\alpha$ for different $\alpha$ with $\gamma=16,\,200$ and $1000$ fitted to $\ssee=a\log\,(n_\mx/\alpha)+b$. The fit parameters are shown in the table which show $a \sim 1/3$. This is also true for other values of $\gamma$. The curves for  complementary values of $\alpha$ are indistinguishable.} \label{allCompl.fig}
\end{figure*}

Fig.~\ref{allCompl.fig} shows the dependence of $\ssee$ on $n_{max}/\alpha$ for different $\alpha$ and with three different values of $\gamma$ $(16, 200\text{ and }1000)$. Here SSEE can be fitted to the form, $\ssee = a_3\log(n_{max}/\alpha) + b_3$. As is clear from the tables in this figure $a_3\approx 0.33\approx 1/3$ for all $\alpha$ and $\gamma$ with the order of error given in the table. $b_3$ however depends on $\alpha$ and $\gamma$. This suggests that $c(\gamma)\approx 1$ in Eqn.~\eqref{ourresult.eq}.

In order to extract $c_1(\gamma)$ we subtract the first term in Eqn.~\eqref{ourresult.eq} (which depends on $n_{max}/\alpha$) using $c(\gamma)/3$ given by the values of $a_3$ in the table of Fig.~\ref{allCompl.fig} from the values of $b_1$ in the table of Fig.~\ref{Allsalpha.fig} for $n_{max}/\alpha = 1200, 2000$ and $2600$. We find that the difference (or $c_1(\gamma)$) is independent of the choice of $n_{max}/\alpha$ which is as expected. We fit the dependence on $\gamma$ by 
\begin{equation}
c_1(\gamma)=a_4\log(\gamma)+b_4
\end{equation}
and the values for the coefficients are given in the table in the Fig.~\ref{conegamma.fig}.

\begin{figure*}[!ht]
	\centering{\subfloat{\includegraphics[height=5cm]{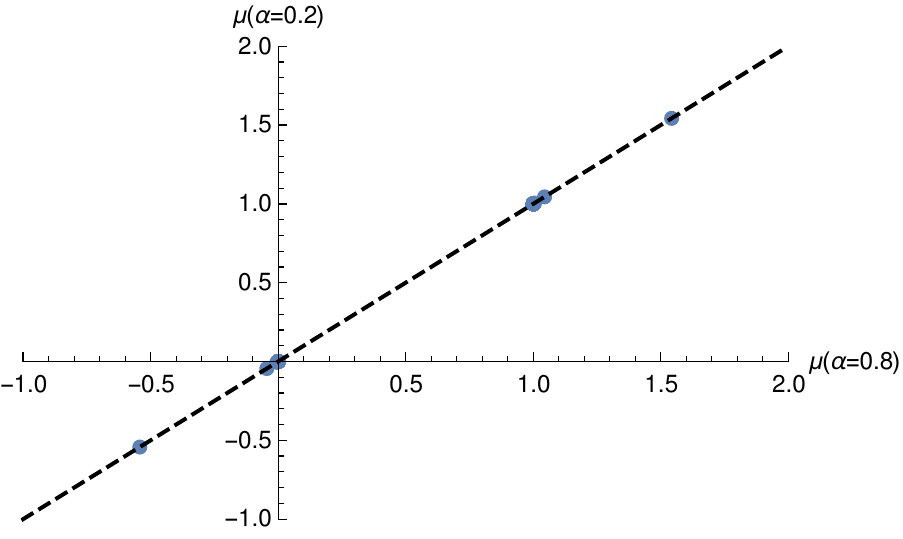}}\,\subfloat{\includegraphics[height=5cm]{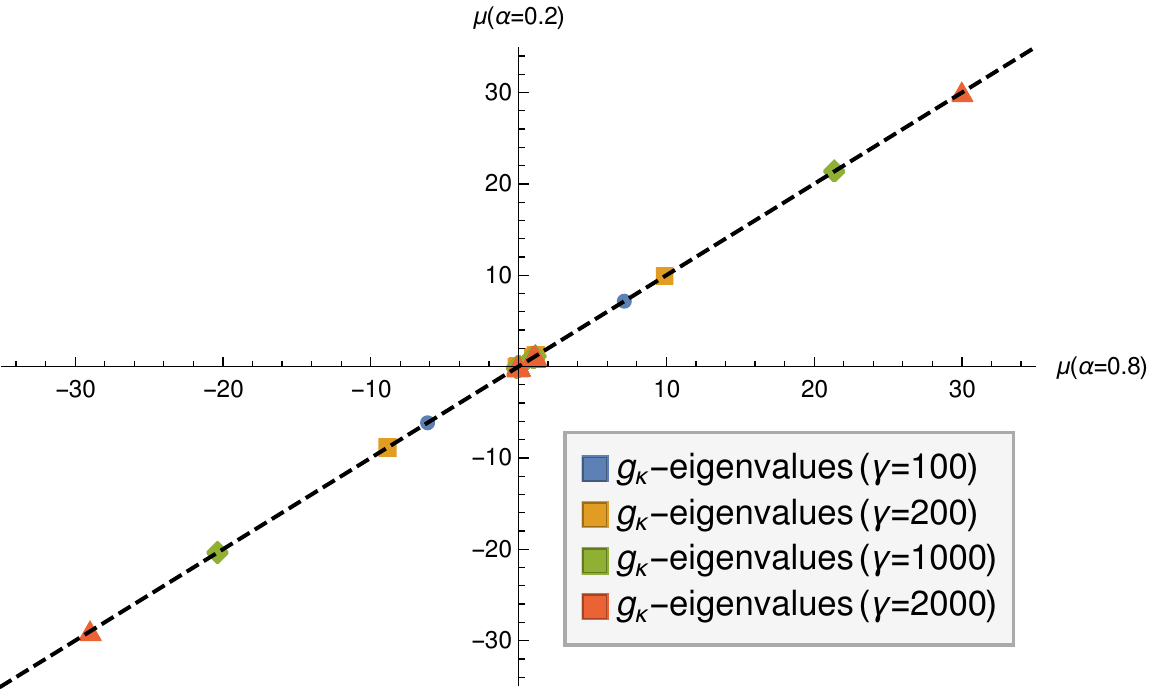}}}
	\caption{A plot comparing the eigenvalues $\mu$ of the entropy equation for one choice of complementary regions
		with $\alpha=0.2,\,0.8$ for $n_\mx/\alpha=2600$ and different choices of $\gamma$. On the left are the eigenvalues
		associated with  the $f_k$ matrix elements which are independent of $\gamma$ and on the right are those associated
		with the $g_\kappa$, which are $
		\gamma$ dependent.  We note that
		the number of eigenvalues differ in both regions but only in the  number of $(1, 0)$ pairs which  leads to the
		equality of the SSEE in these complementary regions. Further, the significant contribution comes from the $g_\kappa$
		matrix elements of which there are precisely {\it two} which are substantially different from $(1,0)$. These increase
		with $\gamma$ and are the main contributors  to $c_1(\gamma)$} 
	\label{ev.fig} 
\end{figure*}

\begin{figure*}[!ht]
	\centering
	\subfloat[$\alpha=0.5$, $n_\mx/\alpha=2600$]{
		\includegraphics[width=5cm]{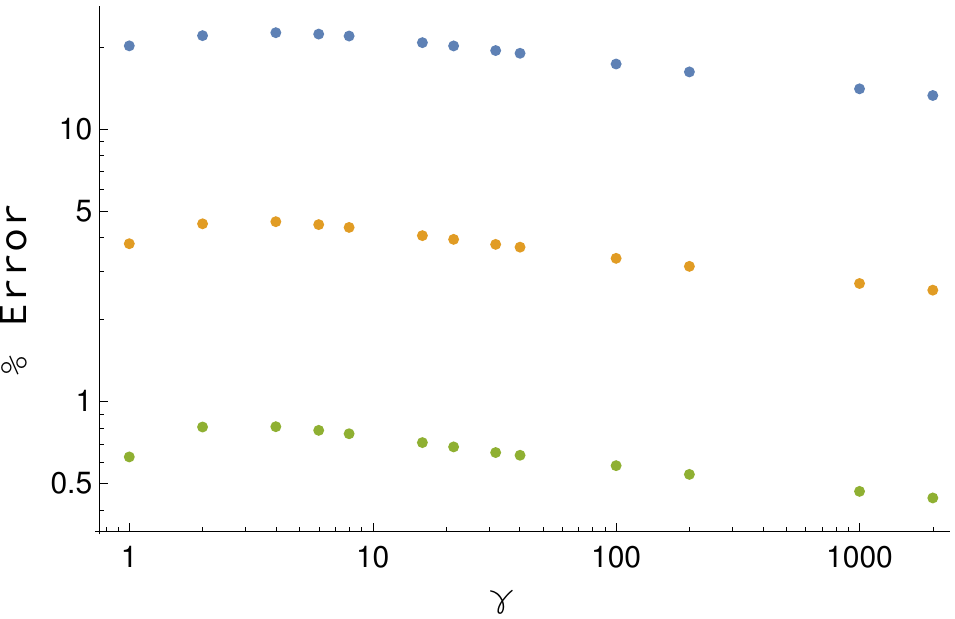} }\, 
	\subfloat[$\alpha=0.5$, $\gamma=1000$]{\includegraphics[width=5cm]{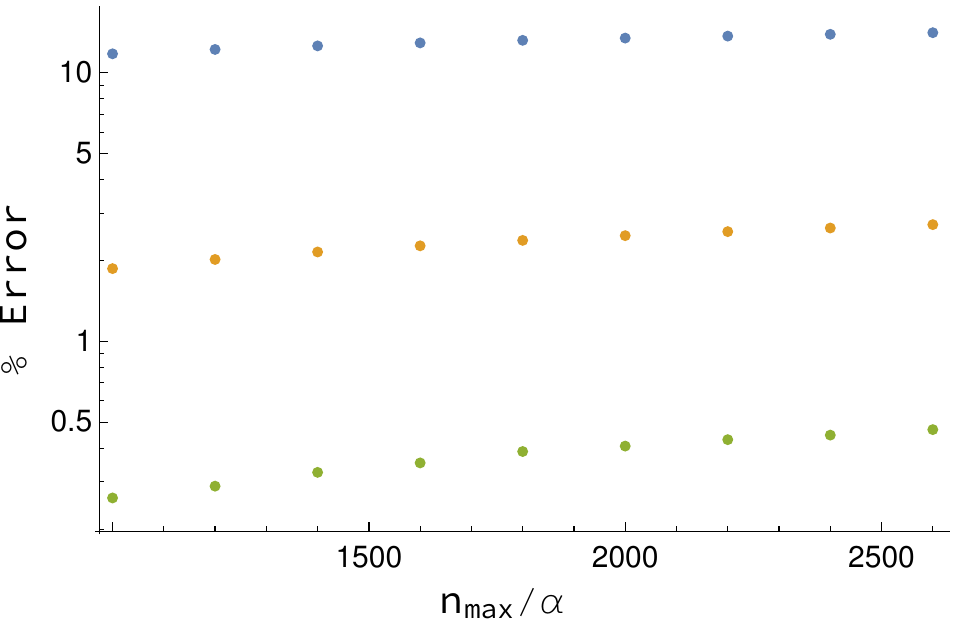}} \, 
	\subfloat[$n_\mx/\alpha=2600$, $\gamma=1000$]{\includegraphics[width=5cm]{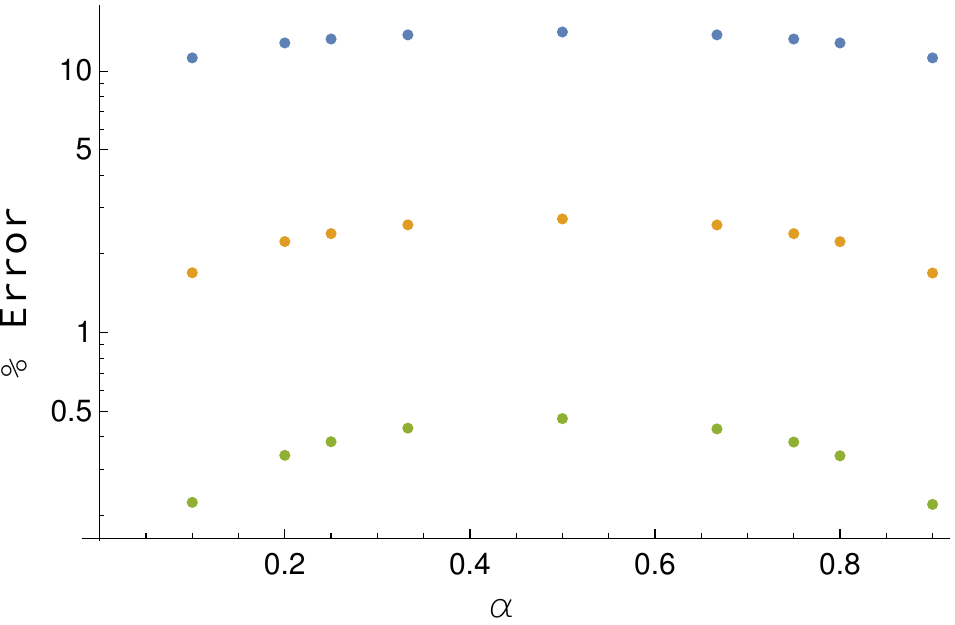}} 
	\caption{In order to estimate the contribution of the pairs $(\mu,1-\mu)$,  we plot the percentage error in the SSEE
		when only the largest pairs (one, two and three represented in blue, orange and green respectively) of eigenvalues are considered, as a function of the different
		parameters $\gamma, n_\mx/\alpha$ and $\alpha$. In each case, we see that the error goes down to $< 1\%$ even when
		only the  3 largest eigenvalues are retained.}
	\label{evtwo.fig} 
\end{figure*}

We also find that the eigenvalues (which always come in pairs $(\mu,1-\mu)$) exhibit the  surprising feature
that all but one pair hovers around the values $0$ and $1$ and
hence contributes significantly to $\ssee$. In Fig.~\ref{ev.fig} we also show the comparison of the eigenvalues obtained in the two complementary regions, we find that they differ only in the numbers of $(0,1)$ pairs. Further, if we calculate $\ssee$ for the largest pairs of eigenvalues, we find that the
error is small, as shown in Fig~\ref{evtwo.fig}.

\bibliography{reference}
\bibliographystyle{ieeetr}
\end{document}